\documentclass[epsfig,psfig,aps,twocolumn,prl]{revtex4-1}
\usepackage[colorlinks=true,citecolor=red,urlcolor=blue]{hyperref}
\usepackage{amsfonts}
\usepackage{graphicx}

\begin{document}
\draft
\title
{
Disorder-induced superconductivity in ropes of carbon nanotubes
}
\author
{
B. Bellafi, S. Haddad and S. Charfi-Kaddour
} 
\affiliation{
Laboratoire de Physique de la Mati\`ere Condens\'ee, D\'epartement de Physique,
Facult\'e des Sciences de Tunis, Campus universitaire 1060 Tunis, Tunisia
}
\date{December 26, 2009}
%
%
\begin{abstract}
We study the interplay between disorder and superconductivity in a rope of metallic carbon nanotubes. Based on the time dependent Ginzburg Landau theory, we derive the superconducting transition temperature T$_c$ taking into account the critical superconducting fluctuations which are expected to be substantially strong in such low dimensional systems. Our results indicate that, contrary to what is expected, T$_c$ increases by increasing the amount of disorder. We argue that this behavior is due to the dynamics of the tubes which reduces the drastic effect of the local disorder on superconductivity by enhancing the intertube Josephson tunneling.
We also found that T$_c$ is enhanced as the effective dimensionality of the rope increases by increasing the number N of the tubes forming the rope. However, T$_c$ tends to saturate for large values of N, expressing the establishment of a bulk three dimensional (3D) superconducting order.

\end{abstract}
\maketitle

\section{Introduction}
Since their discovery in 1991 \cite{Iijima}, carbon nanotubes (CNT) have been studied under close scrutiny due to their eye-catching properties which are of a great interest not only for nanotechnology but also for fundamental physics. A carbon nanotube, which can be regarded as a tiny cylinder rolled up from a graphene sheet, is a good candidate to study electronic properties in one-dimensional (1D) systems where electron-electron interactions are substantially important. \newline
CNT can be synthesized as a single walled tube (SWNT) or multiwalled tubes (MWNT) consisting of two or more concentric shells. SWNT can also be assembled to form ropes of ordered parallel tubes arranged in a triangular lattice \cite{Dressel,Jounet,Thess}. The nanotubes, which are nearly of the same diameter, can have different kind of helicities, but in general $\frac 1 3$ of them are metallic \cite{Ferrier04,Ferrier06}.\newline
The transport properties of the rope is found to be strongly dependent on the amount of disorder within the tubes \cite{Maarouf,Tunney}. It has been reported that the intertube electronic transfer is enhanced in the presence of disorder, leading to a charge carrier delocalization \cite{Ferrier}. This feature raises the question whether such disorder-induced intertube coupling can be observed for the superconducting order in ropes of CNT?\

The first superconducting signature was observed, in 1998, as a proximity effect in isolated metallic bundled SWNT connected to superconducting leads \cite{Kasumov98,Morpurgo}. Later on, intrinsic superconductivity has been reported in ropes of CNT with a transition temperature T$_c=0.55$ K \cite{Ferrier06,Kociak03,Kasumov03}.\
Ferrier {\it et al.} \cite{Ferrier06} studied the dependence of the superconducting transition temperature on the number of the metallic tubes included in the rope and on the amount of disorder. They found that superconductivity arises only in ropes with more than 100 tubes. However, the most striking result of Ref.\cite{Ferrier06} is that disorder, contrary to what is expected, may induce superconductivity: the larger the amount of disorder, the stronger the superconducting correlations. Nevertheless, at a very large disorder amplitude, the superconducting order collapses as in other superconducting materials.\

Superconductivity at T$_c$= 15 K has been also reported in zeolite-inserted SWNT of small diameter (0.4 nm) \cite{Tang}. Takisue {\it et al.} \cite{Takesue} found a superconducting transition at T$_c\sim$ 12 K in MWNT encapsulated in zeolites. These relatively high critical temperatures put the question on the origin of superconductivity in SWNT. How can a superconducting order develop in such low dimensional systems where thermal fluctuations are expected to destroy any long range ordered state?
The surprising observation of superconductivity in CNT has stimulated many theoretical studies to found out the underlying mechanism.\

The realization of a superconducting order in ropes of CNT has been ascribed by Gonzalez \cite{Gonzalez} to the presence of strong attractive electron-electron interactions mediated by phonon exchange. The latter prevails over repulsive Coulomb interaction in ropes with hundred or more of metallic nanotubes.\newline
Other models based on phonon mediated attractive mechanisms have been also proposed\cite{Ferrier04,Sediki,Martino}. In particular, the dependence of the superconducting transition temperature on the number of tubes was quite understood in the framework of the model elaborated by Egger and De Martino \cite{Ferrier04,Martino} who introduced the Josephson couplings between the tubes and the phase fluctuations of the superconducting order parameter. However, a pronounced discrepancy with the experimental data emerges with decreasing the number of the tubes embedded in the rope \cite{Ferrier05,BouchiatP}.\

To explain the relatively high superconducting critical temperature reported in SWNT, Sasaki {\it et al.} \cite{Sasaki} have proposed a new mechanism where superconductivity originates from the edge states specific to graphene. The authors argued that superconductivity is due to a superconductor/normal/superconductor junction where the superconducting phase is realized at the ends of the SWNT while the bulk part of the tube remains metallic.\

An other scenario has been proposed by Zhang {\it et al.} \cite{Zhang} to account for the occurrence of superconductivity in SWNT connected to superconducting or normal electrodes. The authors argued that the SWNT becomes superconducting in the range of 11-30K due to the presence of van Hove singularities in the electron density of states of the nanotube.

Karnaukhov and Diks \cite{Karnau} ruled out the electron-phonon interaction mechanism to explain the formation of the superconducting state in SWNT due to the relatively large value of the critical temperature. The authors suggested an alternative attractive electron-electron interaction originating from strong hybridized interaction induced by the two-band electron structure of SWNT.\

Recently, Belluci {\it et al.} \cite{Belluci} have theoretically argued that superconductivity can arise by a purely electronic mechanism in ultrasmall diameter SWNT and end-bonded multiwalled ones due to the screening of the forward scattering processes.\

More recently, Le Hur {\it et al.} \cite{Lehur} have derived a theoretical model to study the possibility of a superconducting proximity effect in metallic SWNT in the presence of superconducting substrate. The authors showed that the latter induce an unconventional double superconducting gap in the tube.\

The outcome of the above-mentioned studies is that the origin of superconductivity in CNT based systems is still under debate and many relative issues are not yet totally unveiled. In particular, the role of disorder on the stability of the superconducting phase has not been addressed in previous theoretical studies \cite{BouchiatP}. This is a key point which may shed light on the formation of the superconducting phase in low dimensional systems. \

In this paper, we theoretically investigate the effect of disorder on the superconducting state in a rope of CNT. The model is based on the time dependent Ginzburg-Landau theory taking into account the superconducting fluctuations which are substantially important in CNT regarding their low dimensionality. Ferrier {\it et al.} \cite{Ferrier05,FerrierPhD} have actually observed, in ropes of CNT, a large domain of superconducting fluctuations which extends to 1K, namely twice the transition temperature (T$_c$=0.5 K).
In the following we present our model and discuss the obtained results in section III. Section IV is devoted to the concluding remarks.

\section{The model}

We consider a rope of identical SWNT arranged in a triangular lattice characterized by the basis ($\vec{a},\vec{b}$). For simplicity we assume that all the tubes are metallic while experimentally $\frac 2 3$, on average, are semiconductors. This assumption d\oe s not affect the outcomes of the present model which depends basically on the amount of disorder in the rope and on the intertube Josephson couplings. From the numerical point of view, one should expect that our calculated superconducting critical temperatures may be somewhat overestimated compared to the experimental ones since we considered that all the neighboring tubes of a given one are metallic. For a more realistic description, we can consider a random distribution of the tubes with different helicities and diameters. Such complication is, actually, irrelevant for the physics of superconductivity in ropes of CNT since the nature of electronic transport is essentially sensitive to the transverse coupling between the tubes which depends on the intra-tube disorder \cite{Ferrier}.\

The superconducting order is stabilized in the rope via Cooper pair tunneling between tubes and inside a single tube. We denote by $J_1$ and $J_2$ the Josephson coupling parameters across the rope, respectively, to the first and to the second neighboring tubes. We assume that the superconducting phase inside a tube is inhomogeneous with superconducting domains separated by metallic regions. This inhomogeneous structure, which may arise in the presence of impurities, is consistent with the absence of a bulk superconductivity in SWNT \cite{Sasaki}. The superconducting domains along the tube ($z$ axis) are coupled by Josephson tunneling parameterized by $J_0$.\\

Regarding the strong superconducting fluctuations which extend on a large temperature range around the critical temperature T$_c$, the mean field theory breaks down and one should expect clear deviation from the mean field critical temperature T$_0$. These fluctuations can be treated in the frame of the time dependent Ginzburg-Landau (TDGL) theory which has proven to be a reliable tool to study the critical transition region including superconducting fluctuations in different systems such as high-T$_c$ \cite{Puica} and low dimensional organic superconductors \cite{EPL}.\

We start by writing the superconducting free energy $F_s$ of the rope compared to that of the normal state $F_{norm}$: 
\begin{widetext}
\begin{eqnarray}
F&=&F_s-F_{norm}
=\sum_{i,j,n}\int_{r_1}^{r_2}dx\int_{r_1}^{r_2}dy\int_0^{l_{01}}dz \left[a|\psi_{n,i,j}|^2+\frac{{\hbar}^2}{2m^{\ast}}|\vec{\nabla}\psi_{n,i,j}|^2
\right.\nonumber\\
&+& J_0|\psi_{n,i,j}-\psi_{n+1,i,j}|^2
+ J_1|\psi_{n,i,j}-\psi_{\langle n\,i\,j\rangle}|^2
+\left. J_2|\psi_{n,i,j}-\psi_{\langle\langle n,i,j\rangle\rangle}|^2
+\frac b 2 |\psi_{n,i,j}|^4\right],
\label{free}
\end{eqnarray}
\end{widetext}

where $i$ and $j$ denote the tube coordinates in the triangular basis ($\vec{a},\vec{b}$) whereas $n$ indicates the position of the superconducting domain along the tube direction $z$. $\psi_{nij}$ is the superconducting order parameter and $\langle\, \rangle$ and $\langle\langle \; \rangle\rangle$ correspond to the first and second neighboring tubes. The coefficients $a$ and $b$ are given by: $a=a_0\epsilon$ and $b=\mu_0\kappa^2e_0^2{\hbar}^2/2m^2$, where $a_0={\hbar}^2/2m\xi^2_0$, $\xi_0$ being the superconducting coherence length, and $\epsilon=\ln(T/T_0)$ while $\kappa=\frac{\lambda_{\parallel}}{\xi_{\parallel}}$ is the GL parameter. Here $\lambda_{\parallel}$ and $\xi_{\parallel}$ are respectively the London penetration depth and the coherence length in the ($\vec{a},\vec{b}$) plane transverse to the tube direction.
We take for simplicity $\xi_0=\xi_{\parallel}$. The Cooper pair is characterized by its electric charge $e_0=2e$ and its effective mass $m=2 m_e$ where $e$ is the unit charge and $m_e$ is the electron mass.\newline
The superconducting order is assumed to develop inside a tube over a thickness $r_2-r_1$ from the surface. The length of the superconducting domain is denoted $l_{01}$.\newline

The Josephson parameters are written as:

\begin{eqnarray}
&&J_0=\frac{{\hbar}^2}{2m^{\ast} l^2_{02}}\,{\mathrm exp}(-\frac{l_e}L),\quad
J_1=\frac{{\hbar}^2}{2m^{\ast}l^2_1}\,{\mathrm exp}(-\frac{l_e}D), \nonumber\\
&&{ \mathrm and}\quad J_2=\frac{{\hbar}^2}{2m^{\ast}l^2_2}\,{\mathrm exp}(-\frac{l_e}D),
\label{Joseph}
\end{eqnarray}

where $m^{\ast}$ is the effective pair mass in the superconducting domain whereas $l_e$ is the mean free path along the tube. $L$ and $D$ are, respectively the length and the diameter of the rope. We assume for simplicity that all the tubes have the same diameter. $l_1$ ($l_2$) denotes the intertube distance, from the tube surface, between first (second) neighboring tubes while $l_{02}$ is the distance between superconducting domains inside a single tube.\

The natural question which arises concerns the origin of these Josephson coupling expressions. The major issue regards the exponential terms which lead to an enhancement of the Josephson tunneling by increasing the amount of disorder, namely by decreasing the mean free path.\newline
This idea is based on previous studies dealing with Josephson coupled arrays of n-leg spin ladders \cite{Orgad} and correlated stripes in cuprate superconductors \cite{Kivelson} which show clear evidence of the drastic effect of disorder on the superconducting state. 
Kivelson {\it et al.} \cite{Kivelson} have argued that the Josephson coupling between stripes is strongly enhanced by the transverse stripe fluctuations, which promotes the superconducting order. These fluctuations bring neighboring stripes close together leading to the enhancement of the mean value of the Josephson coupling.\

Orgad \cite{Orgad} has shown that such geometrical fluctuations in coupled ladder systems can reduce the suppression of the superconducting correlations due to disorder, by increasing the Josephson tunneling between ladders. The dynamic of the ladders reduces the effective disorder strength and make the superconducting pairing more robust against disorder. The interladder Josephson coupling is found to increase exponentially with the square of the fluctuation amplitude, which enhances the superconducting transition temperature. Orgad \cite{Orgad} considered a Josephson tunneling amplitude depending on the interladder distance as $J_{ij}\sim J_0{\mathrm exp}[-(s+u_i-u_j)/\gamma]$, where $u_i$ and $u_j$ are the deviation of the i$^{th}$ and the j$^{th}$ ladders from their static position, $s$ is the mean distance of the ladder array and $\gamma$ is a characteristic constant \cite{Orgad}.\newline
The basic idea highlighted in Refs.\cite{Orgad,Kivelson} is that the interplay between disorder and the dynamics of the stripes or the ladders is substantial for the stability of the superconducting order in cuprates and spin ladder superconductors.\

Keeping this result in mind, let us now return to the rope of CNT. The latter can be described, as proposed by Ferrier {\it et al.} \cite{Ferrier} by an array of 1D atomic chains lying on a cylinder where each chain corresponds to a SWNT. The hopping processes along the chain are randomly distributed around a mean value $t_{\parallel}$ with a square distribution $\delta t_{\parallel}$. Such bond disorder along the chain may be induced by the dynamics of the tube as in the case of arrays of spin ladder or stripes. This leads to a competition between the geometrical fluctuations of the SWNT and the local disorder inside the tubes.\

By analogy with Ref.\cite{Orgad}, the Josephson tunneling between tubes can be written as $ J\;\alpha \;{\mathrm exp}[-d_{ij}/\gamma]$, where $d_{ij}$ is the separation distance between the i$^{th}$ and the j$^{th}$ tubes. The exponential term expresses the Cooper pair tunneling probability which can be averaged over the tubes as $\langle P_{\perp}\rangle={\mathrm exp}[-\langle d_{\perp}\rangle /\gamma]$, where $ \langle d_{\perp}\rangle$ is an average distance between the tubes. 

In diffusive superconductors, one should expect a dependence of the Josephson couplings on the mean free path since the superconducting coherence length is governed by the disorder amount and reads as $\xi_c=\sqrt{\frac{\hbar v_F l_e}{\Delta}}$, where $\Delta $ is the superconducting gap and $v_F$ is the Fermi velocity \cite{Varlamov}.\
A key question raises at this point concerning the relationship between $ \langle d_{\perp}\rangle$ and the intratube mean free path $l_e$, which we try to answer in the following.\

The plane transverse to the rope direction can be regarded as a dirty two dimensional superconductor of a mesoscopic size where the disorder points, due to defects or impurities, are localized inside the tubes. In this plane, the tube sections form a sort of disordered clusters embedded in a free disorder medium. The average distance $ \langle d_{\perp}\rangle$ between these clusters is controlled by the dynamic of the tube which is strongly dependent on the disorder amount inside the tubes.
In the diffusive regime, the bond disorder due to the geometrical fluctuations of the tubes gives rise to an increasing intertube one particle hopping integral with increasing the site disorder amplitude originating from impurities and defects inside the tube \cite{Ferrier}. This means that the intertube distance $ \langle d_{\perp}\rangle$ decreases with decreasing the intratube mean free path $l_e$. $ \langle d_{\perp}\rangle$ is then expected to have the same behavior as $l_e$ and may be expressed as a growing function of $l_e$.
We do not claim that the present model provides the exact form of this function. A more detailed analysis based on a microscopic study is needed. 

Since $ \langle d_{\perp}\rangle$, as  $ l_e$, is a free parameter in our model, we set for simplicity $ \langle d_{\perp}\rangle =l_e$. This means that, in the diffusive regime, the mean free path inside the tube and across the rope are of the same order. This is justified as far as $ l_e$ is smaller than the rope diameter $D$ to keep the transverse one particle transport in the diffusive regime. Actually, this approximation does not affect the overall outcomes of our model but may yields to somewhat larger superconducting critical temperatures compared to the experimental ones.

To characterize the electronic transport in disordered mesoscopic systems, one need to compare the size of the system, which is the rope diameter in this case, to a characteristic mean free path.
Regarding its dependence on the intratube disorder amplitude, $ \langle d_{\perp}\rangle$ seems to be a good parameter to account for the transport regime across the rope.
It comes out that $ \langle d_{\perp}\rangle$ and the rope diameter $D$, which depends on the tube number N, are the key parameters for the one particle transport and for the Cooper pair tunneling across the rope in the diffusive regime. 
The tunneling probability can then be written as $\langle P_{\perp}\rangle={\mathrm exp}[-\langle d_{\perp}\rangle /\gamma]= {\mathrm exp}[-l_e/D]$, where the $\gamma$ constant, which accounts for the environment between the tubes, is replaced by rope diameter $D$. This is made possible since the tube environment is disorder free and depends only on the tube number included in the expression of the rope diameter $D$.\

In the absence of site disorder and geometrical fluctuations, namely in a pure static rope, the Josephson couplings between respectively the first and the second neighboring tubes write as:

\begin{eqnarray}
J_1=\frac{{\hbar}^2}{2m^{\ast}l^2_1}\quad{ \mathrm and}\quad J_2=\frac{{\hbar}^2}{2m^{\ast}l^2_2}
\label{Joseph2}
\end{eqnarray}
Such couplings cannot describe the superconducting order in the rope since they are independent on the rope characteristics particularly the tube number.\newline
In the presence of disorder and geometrical fluctuations of the tubes, the Josephson parameters $J_1$ and $J_2$ given by Eq.\ref{Joseph2} should be changed to account for the average pair tunneling probability across the rope $\langle P_{\perp}\rangle={\mathrm exp}[-l_e/D]$, which gives rise to the expressions introduced in Eq.\ref{Joseph}.\

Regarding the intratube Josephson tunneling $J_0$, one can define an average pair hopping probability along the tube $\langle P_{\parallel}\rangle={\mathrm exp}[-l_e/L]$ resulting from the geometrical fluctuations of the tube which yields to the expression given by Eq.\ref{Joseph}.\newline
It is worth to note that the $J_0$ term is irrelevant for the stability of the superconducting phase as we will show in the next.\

It comes out that the dynamics of the tubes in the rope mitigate the drastic effect of the local disorder on the superconducting order by enhancing the Josephson tunneling amplitudes between the tubes. The latter increase as a function of the effective disorder. This is reminiscent of the disorder-induced electronic transverse delocalization in ropes of CNT proposed by Ferrier {\it et al.} \cite{Ferrier}. We suggest that this delocalization scenario holds for Cooper pair due to the tube dynamics as argued above.\\

Let us now turn to the superconducting order parameter whose critical dynamics satisfy the TDGL equation:
\begin{eqnarray}
\Gamma^{-1}_0\frac{\partial \psi_{nij}}{\partial t}=-
\frac{\partial F}{\partial\psi^{\ast}_{nij}}+\zeta_{nij}(\vec{r},t)
\end{eqnarray}
Here $\Gamma^{-1}_0=\pi{\hbar}^3/16 m \xi^2_{\parallel} k_BT$ is the relaxation rate of the order parameter whereas $\zeta_{nij}(\vec{r},t)$ are the Langevin forces describing the thermodynamical fluctuations and which obey the Gaussian white-noise law\cite{Puica}:
\begin{eqnarray*}
\langle \zeta_{nij}(\vec{r},t) \zeta^{\ast}_{n^{\prime}i^{\prime}j^{\prime}}(\vec{r}\;^{\prime},t^{\prime})
\rangle=2
\Gamma^{-1}_0k_BT\delta(\vec{r}-\vec{r}\;^{\prime})\delta(t-t^{\prime})
\end{eqnarray*}
with $\vec{r}=(X+id,Y+jd,Z+nl_0)$
and $\vec{r}\;^{\prime}=(X+i^{\prime}d,Y+j^{\prime}d,Z+n^{\prime}l_0)$, where $d=l_1+d_0$ and $l_0=l_{01}+l_{02}$, $d_0$ being the tube diameter. X, Y and Z are the coordinates of a point belonging to a superconducting domain of a SWNT of length $l_{01}$, along the $z$ direction, and of a thickness $r_2-r_1$.\

By taking the derivative of the free energy (Eq.\ref{free}) with respect to $\psi^{\ast}_{nij}$, the TDGL equation becomes:
\begin{widetext}
\begin{eqnarray}
&&\zeta_{nij}(\vec{r},t)=
\Gamma^{-1}_0 \frac{\partial \psi_{n,i,j}}{\partial t}+ a\psi_{n,i,j}-\frac{{\hbar}^2}{2m^{\ast}} \Delta \psi_{n,i,j}
+b\langle|\psi_{n,i,j}^2|\rangle \psi_{n,i,j}
+6\,J_1\psi_{n,i,j}\nonumber\\
&-&J_1\left( \psi_{n,i+1,j}+\psi_{n,i-1,j}
+\psi_{n,i,j+1}+\psi_{n,i,j-1}+\psi_{n,i+1,j-1}+\psi_{n,i-1,j+1}\right)\nonumber\\ 
&+& J_2\left(6\psi_{n,i,j}-\psi_{n,i+2,j-1}-\psi_{n,i-2,j+1}
-\psi_{n,i+1,j+1}-\psi_{n,i-1,j-1}\right) \nonumber\\
&-&
J_2\left(\psi_{n,i+1,j-2}+\psi_{n,i-1,j+2}\right)
+J_0\left( 2\psi_{n,i,j}-\psi_{n+1,i,j}-\psi_{n-1,i,j}\right)
\label{zeta}
\end{eqnarray}
\end{widetext}
where we adopted the Hartree approximation for the quartic term as in Ref.\onlinecite{Puica}, which results in replacing the term $b|\psi_{nij}|^2\psi_{nij}$ by $b\langle|\psi_{nij}|^2\rangle \psi_{nij}$. This approximation leads to a linear problem with a reduced temperature:
\begin{equation}
\tilde{\epsilon}=\epsilon+\frac b a \langle|\psi_{nij}|^2\rangle,
\label{self}
\end{equation}
which is determined self-consistently together with 
$\langle|\psi_{nij}|^2\rangle$. The superconducting critical temperature is defined as $\tilde{\epsilon}(T=T_c)=0$\cite{Puica}.\newline
To solve this equation, we introduce the Fourier transform of $\psi_{nij}$ as

\begin{eqnarray*}
\psi_{nij}(\vec{r},t)&=&\int\frac{d^3\vec{k}}{(2\pi)^3}\psi(\vec{k},t)
{\rm e}^{-i\vec{k}. \vec{r}}
\end{eqnarray*}
where

\begin{eqnarray*}
&&\psi(\vec{k},t)=\sum_{nij}\int_{X_1}^{X_2}dX\int_{Y_1}^{Y_2}dY \int_{0}^{l_{01}}dZ \times\nonumber\\
&&\psi_{nij}(X+id,Y+jd,Z+nl_0,t)
{\rm e}^{(X+id)k_x}\;
{\rm e}^{(Y+jd)k_y}\;{\rm e}^{(Z+nl_0)k_z},
\end{eqnarray*}
where $X_1$ and $X_2$ ($Y_1$ and $Y_2$) are the limiting values for $X$ ($Y$) in a superconducting domain in the ($a,b$) plane.

Taking the Fourier transform of Eq.\ref{zeta}, we obtain
\begin{widetext}
\begin{eqnarray}
\zeta(\vec{k},t)=
&&\left\{\Gamma^{-1}_0 \frac{\partial}{\partial t}+ \frac{{\hbar}^2k^2}{2m^{\ast}} 
+\tilde{a}+2J_0\left( 1-\cos(k_zl_0)\right)
+2J_1\left[3-\cos(dk_x)-\cos(dk_y)-\cos\left(d(k_x-k_y)\right)\right]\right.\nonumber\\
&+&\left.2J_2\left[3-\cos\left(d(2k_x-k_y)\right)-\cos\left(d(k_x+k_y)\right)-\cos\left(d(k_x-2k_y)\right)\right]\right\}\psi(\vec{k},t),
\label{zetak}
\end{eqnarray}
\end{widetext}
with $\tilde{a}=a+b\langle|\psi_{nij}|^2\rangle$ and the correlation relation satisfied by $\zeta(\vec{k},t)$ :
\[
\langle \zeta(\vec{k},t)\zeta^{\ast}(\vec{k}^{\prime},t^{\prime})\rangle=2\Gamma^{-1}_0k_BT(2\pi)^3\delta(\vec{k}-\vec{k}^{\prime})\delta(t-t^{\prime})
\]
Equation \ref{zetak} can be solved using the Green function method proposed by Puica and Lang \cite{Puica} for layered superconductors. We define the Green function $R(\vec{k},t,k^{\prime}_z,t^{\prime})$ through the relation:

\begin{eqnarray}
&&\left[\Gamma^{-1}_0 \frac{\partial}{\partial t}+ \frac{{\hbar}^2k^2_z}{2m^{\ast}} 
+2J_0\left(1-\cos(k_zl_0)\right)+a_1\right]\times\nonumber\\
&&R(\vec{k},t,k^{\prime}_z,t^{\prime})=\delta(k_z-k^{\prime}_z)\delta(t-t^{\prime}),
\label{Rk}
\end{eqnarray}

where 
\begin{widetext}
\begin{eqnarray}
a_1&=&\tilde{a}+\frac{{\hbar}^2(k^2_x+k^2_y)}{2m^{\ast}}
+2J_1\left[3-\cos(dk_x)-\cos(dk_y)-\cos\left(d(k_x-k_y)\right)\right]\nonumber\\
&+&2J_2\left[3-\cos\left(d(2k_x-k_y)\right)-\cos\left(d(k_x+k_y)\right)-\cos\left(d(k_x-2k_y)\right)\right].
\end{eqnarray}
\end{widetext}
We also introduce the Fourier transform of $R(\vec{k},t,k^{\prime}_z,t^{\prime})$ with respect to time as:
\begin{eqnarray}
R(\vec{k},\omega,k^{\prime}_z,t^{\prime})=\int dt R(\vec{k},t,k^{\prime}_z,t^{\prime})
{\rm e}^{i\omega (t-t ^{\prime})},
\end{eqnarray}
which can be deduced from Eq.\ref{Rk} as:

\begin{eqnarray}
&&R(\vec{k},\omega,k^{\prime}_z,t^{\prime})=\delta(k_z-k^{\prime}_z)\times\nonumber\\
&&\left[-i\omega \Gamma^{-1}_0 + \frac{{\hbar}^2k^2_z}{2m^{\ast}} 
+2J_0\left(1-\cos(k_zl_0)\right)+a_1\right]^{-1},
\label{TFRk}
\end{eqnarray}

$\psi(\vec{k},t)$ solution of Eq.\ref{zetak} can be expressed in term of the Green function $R(\vec{k},t,k^{\prime}_z,t^{\prime})$ as \cite{Puica}
\[
 \psi(\vec{k},t)=\int dt^{\prime}\int dk^{\prime}_z R(\vec{k},t,k^{\prime}_z,t^{\prime})
\zeta(k_x,k_y,k^{\prime}_z,t^{\prime}).
\]
Given Eq.\ref{TFRk}, we obtain:

\begin{eqnarray}
&&\psi(\vec{k},t)=\int_0^{\infty}d\tau \zeta(\vec{k},t-\tau)\int d\omega {\rm e}^{i\omega \tau}\times\nonumber\\
&&\left[-i\omega \Gamma^{-1}_0 + \frac{{\hbar}^2k^2_z}{2m^{\ast}} 
+2J_0\left(1-\cos(k_zl_0)\right)+a_1\right]^{-1}
\end{eqnarray}

with $\tau=t-t^{\prime}$ and the following correlation relation:
\begin{eqnarray}
\langle\psi(\vec{k},t)\psi^{\ast}(\vec{k}^{\prime},t)\rangle\; \alpha\;\delta(\vec{k}-\vec{k}^{\prime})
\label{correlation}
\end{eqnarray}
To solve Eq.\ref{self}, one need to derive $\langle|\psi_{nij}|^2\rangle$ which, regarding Eq.\ref{correlation}, can be simply written as:

\begin{eqnarray}
&&\langle|\psi_{nij}|^2\rangle=4\pi \Gamma^{-1}_0k_BT\int d\omega\int d^3\vec{k}\times\nonumber\\
&&\left\{(\omega \Gamma^{-1}_0 )^2+ \left[\frac{{\hbar}^2k^2_z}{2m^{\ast}} 
+2J_0\left(1-\cos(k_zl_0)\right)+a_1\right]^{2}\right \}^{-1}
\end{eqnarray}

The critical temperature can now be deduced from Eq.\ref{self} by setting $\tilde{\epsilon}(T=T_c)=0$, which yields to Eq.\ref{Tc} given in the Appendix.
In the next we discuss the numerical results.

\section{Results and discussion}
We have solved numerically Eq.\ref{Tc} and the results are depicted in Fig.1 which shows the superconducting transition temperature T$_c$ as a function of the number $N$ of the tubes forming the rope. It is worth to note that $N$ is involved in the rope diameter $D$ as $D=\sqrt{N}(d_0+e)$ where $d_0$ and $e$ are respectively the tube diameter and the intertube distance \cite{Kasumov03}.\

As shown in Fig.1, T$_c$ is strongly enhanced by increasing $N$ but this enhancement is slowed down for $N$ larger than 100 with a tendancy to saturation, which is reminiscent of the experimental results \cite{Ferrier04,FerrierPhD}. This behavior reflects the dimensionality of the superconducting phase appearing in the rope. By increasing $N$, the 3D character of the superconducting state is enhanced and T$_c$ likewise. However, for a larger $N$ ($N\sim 200$), the rope can be regarded as a 3D system and a further increase of $N$ is irrelevant for the superconducting order, which explains the saturation behavior of T$_c$ at large $N$.\

%
\begin{figure}[t] 
\begin{center}
\includegraphics[width=7cm,height=5cm]{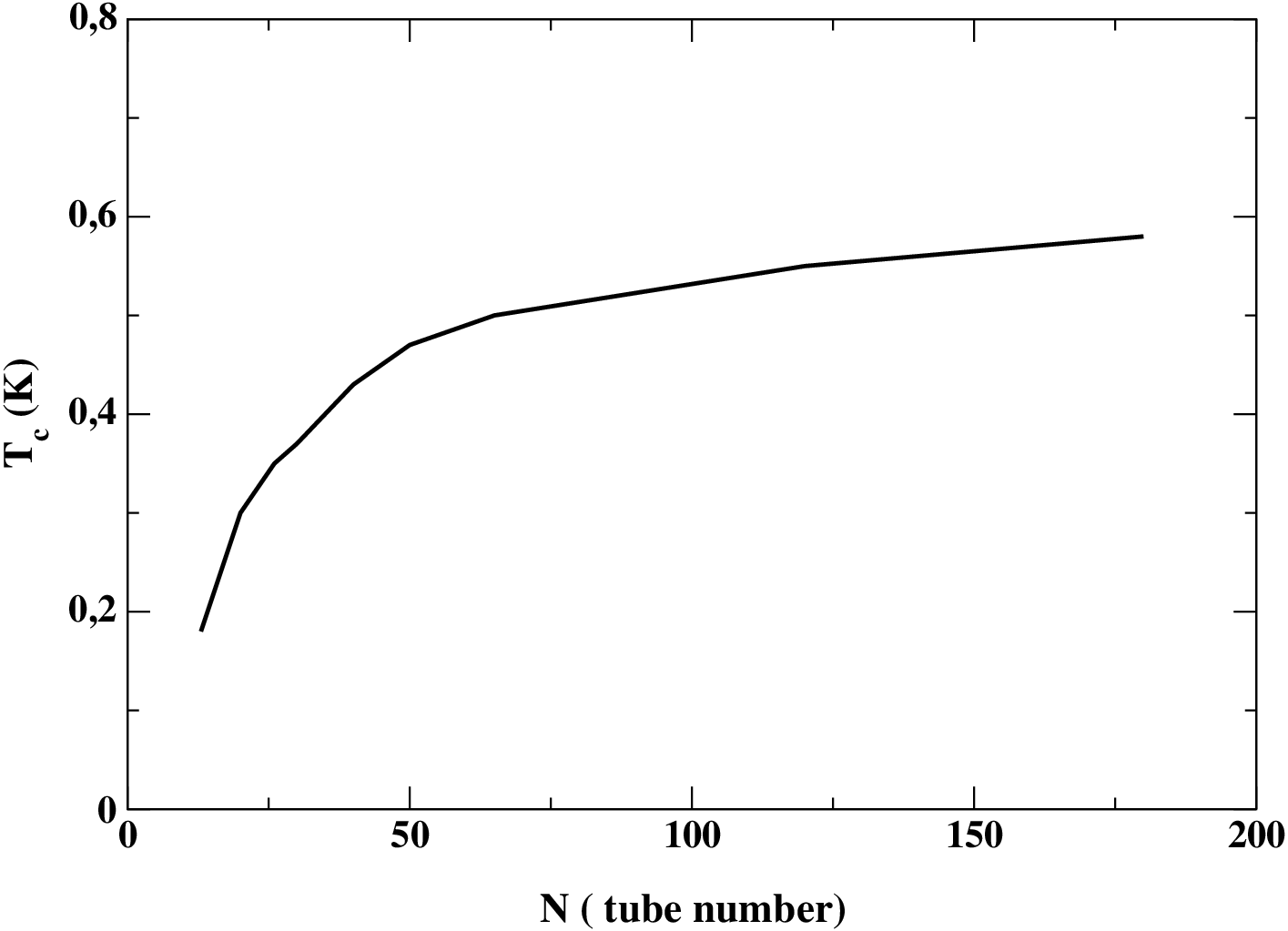}
\end{center}
\caption{Superconducting transition temperature as a function of the number $N$ of tubes. The calculations are done in the one-particle delocalized regime and for $\lambda=0.6 \mu$m, $\xi_0=0.1\mu$m and $L=1.4\mu$m \cite{Ferrier05}. $\lambda$ and $\xi_0$ are respectively the penetration depth and the coherence length in the superconducting domain while $L$ is the rope length.} 
\label{fig1}
\end{figure}

A worth noting question concerns the interplay between superconductivity and the 1D character of a SWNT. Could superconductivity prevail over the low dimensionality of such systems? This turns out to consider only the $J_0$ term in our model. In such case, numerical calculations show that T$_c$ is at most of the order of 1 mK, which explains the difficulty to observe an intrinsic superconductivity in SWNT as reported in Refs.\onlinecite{Ferrier,Ferrier05}. The superconducting phase can, actually, develop in ropes containing about one hundred metallic tubes as shown by earlier studies \cite{Gonzalez,Martino}. The limiting tubes number in our model is then N=13 if one include the first and second neighbors of a given tube.\

In Fig.2, we give, for different tube numbers, the dependence of T$_c$ on the inverse of the mean free path which mimics the amount of local disorder inside the tube. Peculiarly, Fig.2 shows that disorder promotes the superconducting order as found experimentally \cite{Ferrier,Ferrier05}. This behavior is due to the intertube disorder-induced delocalization of the Cooper pairs. Actually, the intertube pair delocalization is expected to develop in the electronic diffusive regime, where disorder can induce transverse hopping processes across the rope \cite{Ferrier}. \

It is worth to note that the values of the critical temperature reported in Figs.1 and 2 may be somewhat overestimated since we have considered that all the tubes are metallic. In a more realistic model, one should take, on average, for each tube two neighboring metallic tubes since, in most cases, $\frac 1 3$ of the tubes within a rope are metallic.
%
\begin{figure}[t] 
\begin{center}
\includegraphics[width=7cm,height=5cm]{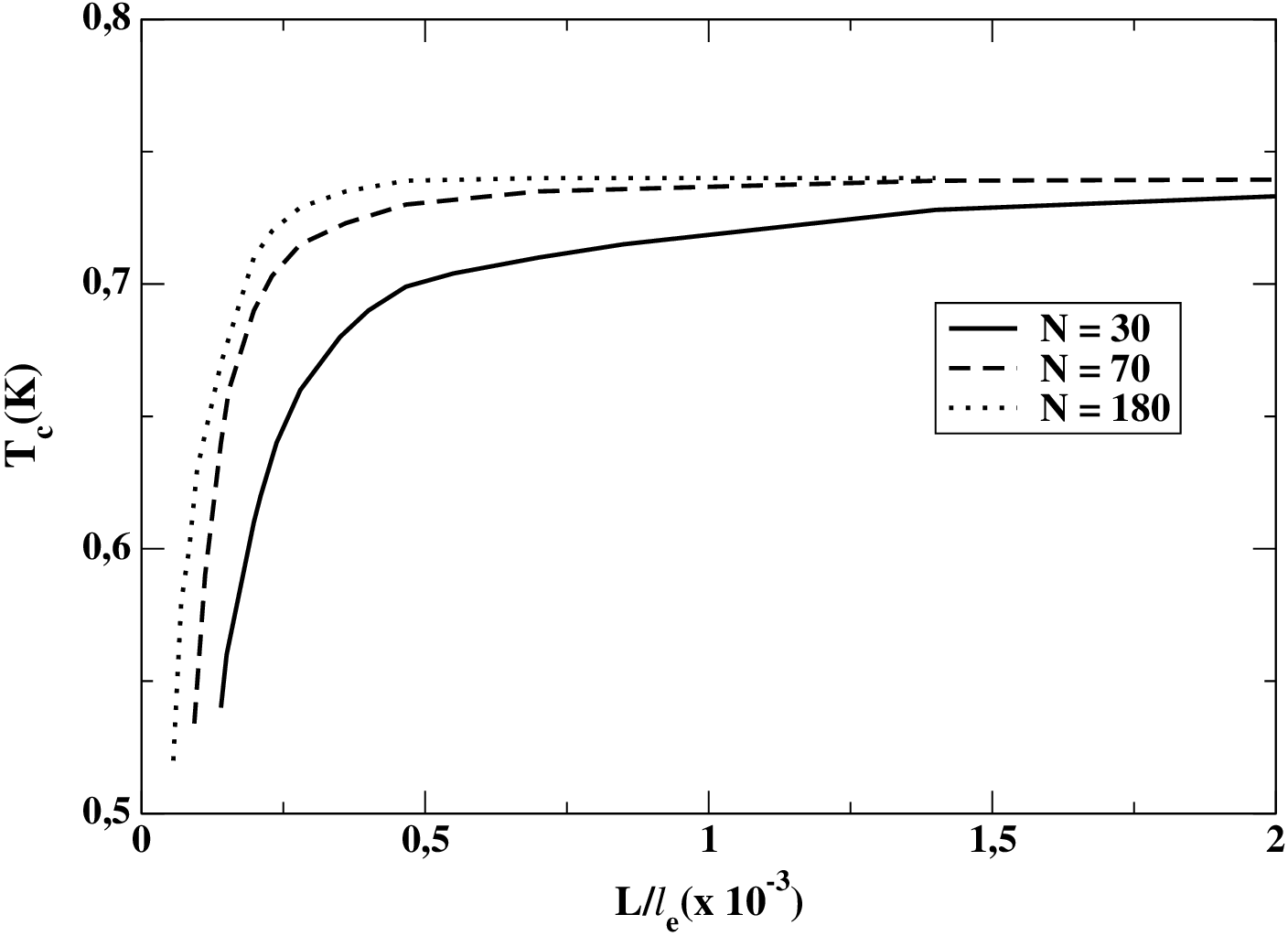}
\end{center}
\caption{Superconducting transition temperature as a function of the inverse of the mean free path in a rope of SWNT for different tube numbers. The calculations are done in the one-particle delocalized regime and for the same data as in Fig.1} 
\label{fig2}
\end{figure}

By the way, one should emphasize the role of the Josephson tunneling $J_2$ between second neighboring tubes on the superconducting order. Numerical results show that $T_c$ is reduced by  $20\%$ if $J_2$ is neglected. Actually, the second neighboring tubes should be involved in the tunneling processes since they are in the same range of reach as the first neighbors \cite{BouchiatP}. This is due to the geometry of the rope characterized by a tube diameter of 3 nm and an intertube distance of 0.35 nm.\\

A worth stressing question regards the saturation behavior of $T_c$ at large disorder amplitude in Fig.2. This feature, which is due to the expression of the intertube couplings given by Eq.\ref{Joseph}, does not sound in agreement with the experimental data which rather show a collapse of the superconducting phase at large enough amount of disorder\cite{Kasumov03,FerrierPhD}. This discrepancy originates from the nature of the electronic transport regime. Our results are derived within the delocalized diffusive regime characterized by a disorder induced transverse electronic hopping \cite{Ferrier}. However, in the large disorder range, a localized regime develops where the electrons are confined within individual tubes. The Josephson couplings given by Eq.\ref{Joseph} are no more reliable since, in this case, the intratube disorder overcomes the geometrical fluctuations of the tubes, leading to the suppression of the intertube pair tunneling. The latter is expected to be strongly reduced by the electron localization which can be roughly described by an exp$\left(-\frac{L}{\xi}\right)$ behavior for the intertube electron hopping where $L$ is the rope length and $\xi=2Nl_e$ is the localization length \cite{Ferrier}. $N$ and $l_e$ being the number of metallic tubes and the mean free path inside the tube. As a consequence, one can assume the following Josephson couplings:

\begin{eqnarray}
J_1=\frac{{\hbar}^2}{2m^{\ast}l^2_1}\,{\mathrm exp}(-\frac{L}{\xi}) \,
{ \mathrm and}\quad J_2=\frac{{\hbar}^2}{2m^{\ast}l^2_2}\,{\mathrm exp}(-\frac{L}{\xi}),
\label{Joseph3}
\end{eqnarray}
which express the disorder induced Cooper pair localization as a result of the electronic localization.\

Fig.3 shows the superconducting transition temperature $T_c$ as a function of the inverse of the mean free path $l_e$ which is a measure of the disorder amplitude. The calculations are done using Eq.\ref{Joseph3}. In this regime of localization, $T_c$ is reduced by increasing disorder due to the suppression of the intertube tunneling. However, the tube number $N$ acts, as in the delocalized regime, to the benefit of the superconducting phase. Increasing $N$ furthers the establishment of a 3D electronic transport regime by increasing the localization length $\xi$. The effect of disorder is significantly important in ropes with a small tubes number where the 1D character prevails over the formation of a 3D superconducting order.\

%
\begin{figure}[t] 
\begin{center}
\includegraphics[width=7cm,height=5cm]{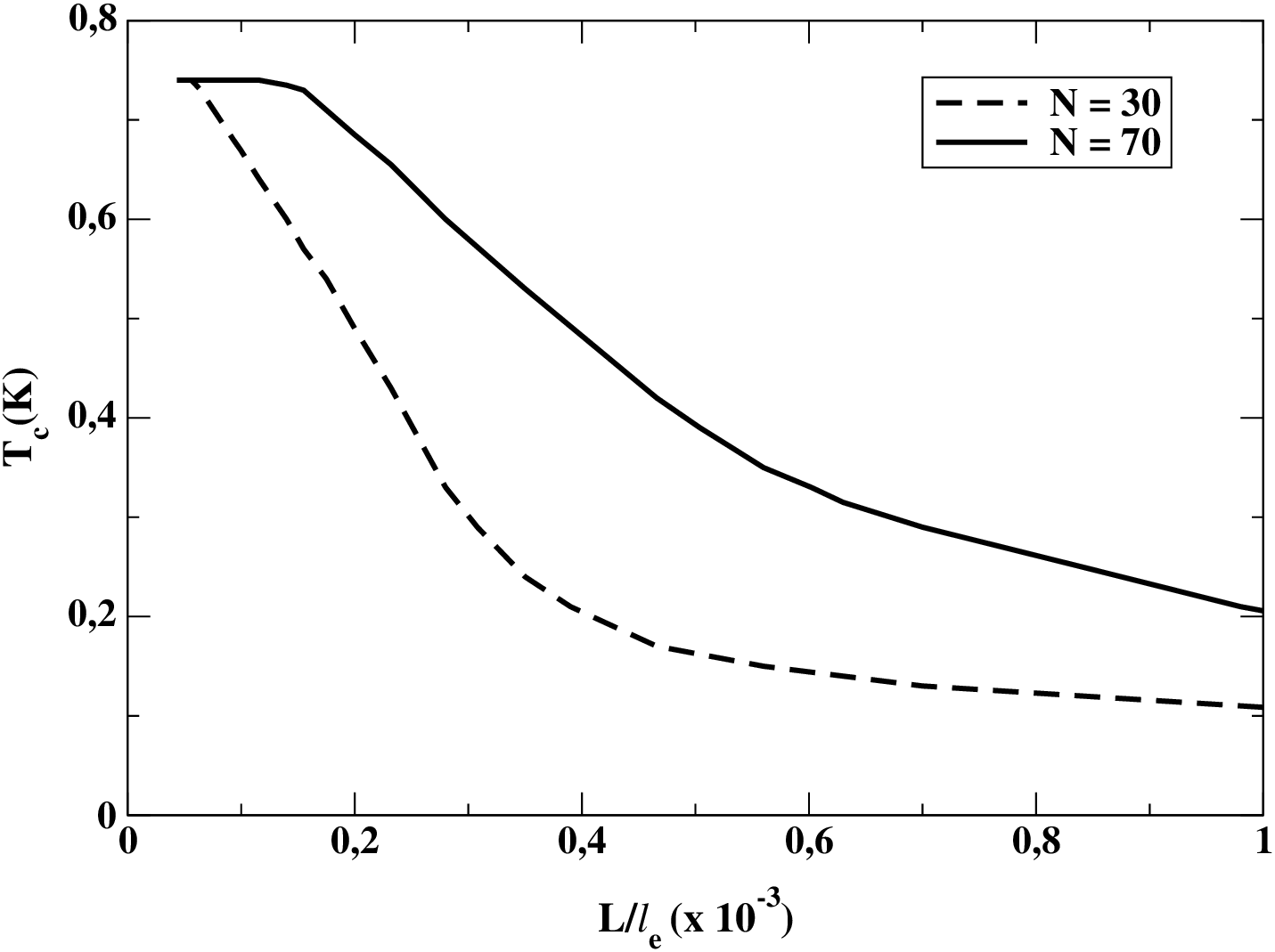}
\end{center}
\caption{Superconducting transition temperature as a function of the inverse of the mean free path in ropes of N=30 and N=70 tubes. $T_c$ is calculated in the localization regime where the Josephson couplings are given by Eq.\ref{Joseph3}. The used data are the same as in Figs.1 and 2.} 
\label{fig3}
\end{figure}

The superconducting behaviors in the delocalized and localized regimes (Figs.2 and 3), are reminiscent of those obtained in a 2D array of stripes \cite{Kivelson}. In such systems, the superconducting transition temperature is found to increase with the transverse stripe fluctuations up to a critical value above which it drops. This happens when the system undergoes a phase transition to an isotropic state where the stripe structure is lost.\\

A tough question raised from Figs.2 and 3 concerns the extension of the delocalized regime. At which disorder amplitude the dynamic of the tubes is frozen and the intertube Josephson tunnelings start to collapse? A rough estimation may be deduced from the experimental results of Kasumov {\it et al.} \cite{Kasumov03} showing that the key parameter governing the disorder in a suspended rope is the ratio $\frac{\xi_c}L$ where $L$ and $\xi_c$ are respectively the rope and the coherence lengths. The latter depends on the mean free path $l_e$ as discussed above $\xi_c=\sqrt{\frac{\hbar v_Fl_e}{\Delta}}$\cite{Varlamov}.\

In Fig.4, we have depicted the behavior of superconducting transition temperature in the localized and delocalized regimes for a rope of N=70 tubes based on the results shown in Figs.2 and 3. According to Fig.4, the suppression of the superconducting order starts at a critical value $\frac{L}{l_{ec}}=$0.2. The smaller the tube number, the greater $l_{ec}$, the frailer the superconducting order.\

%
\begin{figure}[t] 
\begin{center}
\includegraphics[width=7cm,height=5cm]{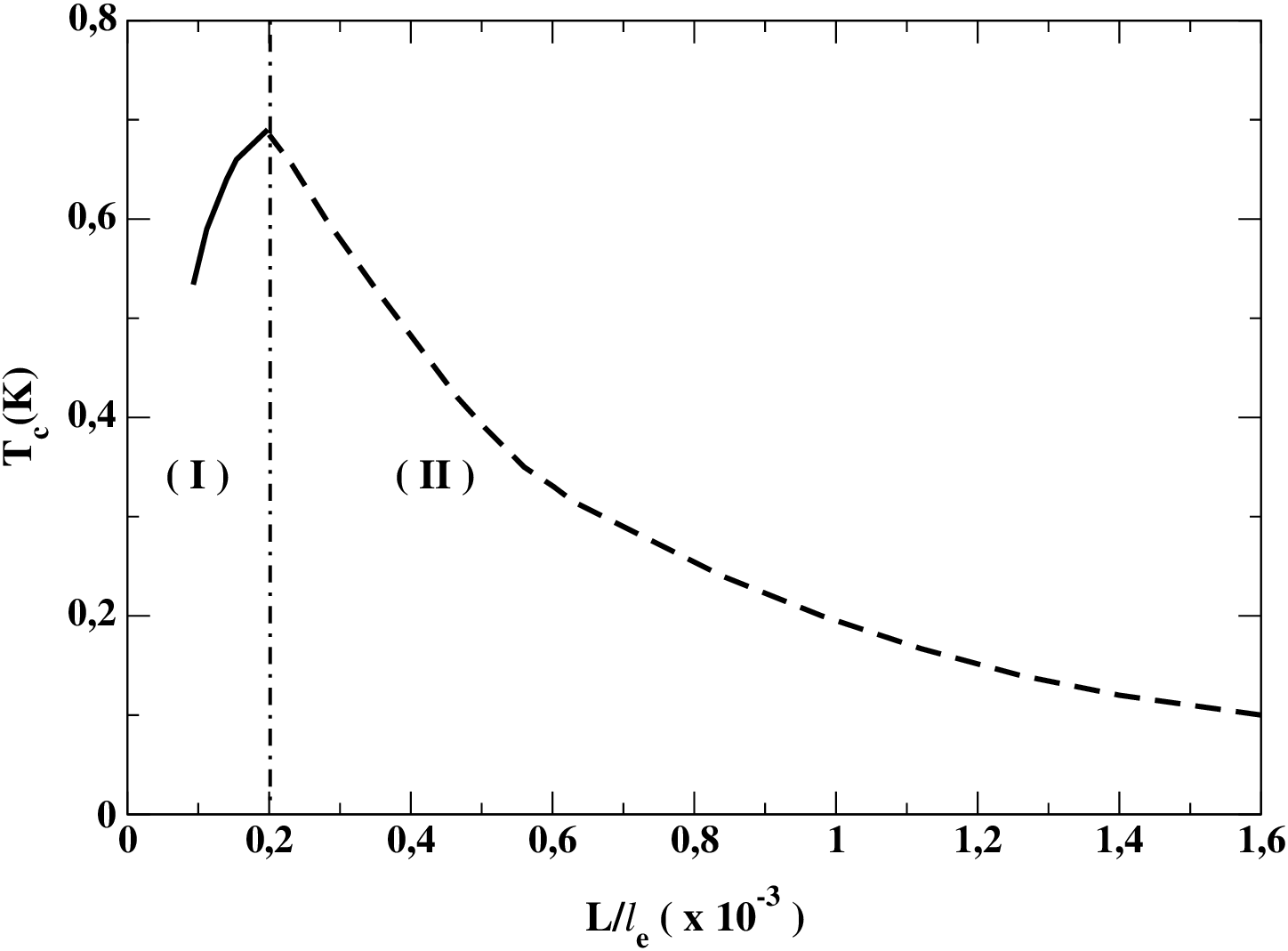}
\end{center}
\caption{Superconducting transition temperature as a function of the inverse of the mean free path in ropes of N=70 tubes. Regions (I) and (II) denote, respectively, the delocalized and localized regimes. The calculations are done with the same data as in Fig.1} 
\label{fig4}
\end{figure}

The comparison of the numerical values of $l_{ec}$ with the experimental results of Ref.\onlinecite{Ferrier} is not obvious. More data are needed to accurately determine the critical disorder amplitude at which the superconducting transition temperature reaches its maximum before decreasing. 
Nevertheless, one can compare the extent of the disorder regime over which the superconducting order develops. Let us characterize this disorder range by the ratio $\eta=\frac{l_{e1}} {l_{e2}}$, where $l_{e1}$ and $l_{e2}$ are respectively the mean free paths corresponding to the appearance and the collapse of the superconducting phase.\
According to the data of Ferrier {\it et al.}\cite{Ferrier,FerrierPhD}, superconductivity appears at $\left(\frac{\xi_c}L\right)_1=\frac 12$ and vanishes at $\left(\frac{\xi_c}L\right)_2=\frac 1{10}$. Assuming that $\xi_c \alpha \sqrt{l_e}$ \cite{Ferrier,FerrierPhD} and a constant rope length $L$, gives rise to $\eta=\frac{\xi_{c1}^2}{\xi_{c2}^2}=\frac{l_{e1}} {l_{e2}}=25$.
From Fig.4, $\eta=\frac{1.6}{0.09}\sim 18$, where we consider that $T_c=1$mK corresponds to the disappearance of the superconducting phase. This value is quite in agreement with the experimental one.
Moreover, one can estimate from Fig.4 the range of the disorder-induced superconductivity regime to which, one may assign a ratio $\eta_d=\frac{l_{e1}}{l_{ec}} \sim 2.2$, namely $\frac 17$ of the total disorder regime over which superconductivity may be observed. Checking this value requires more experimental data.

\section{Concluding remarks} In summary using TDGL theory, we probed the role of the effective dimensionality and the amount of disorder on the stability of the superconducting order in ropes of CNT. We found that an increase of the dimensionality of the rope, which is achieved by increasing the tube number $N$, promotes the establishment of a 3D superconducting phase with an increasing superconducting critical temperature $T_c$. However, for large $N$ values, $T_c$ tends to saturation indicating the formation of a well defined 3D superconducting order.\
The main result of our work regards the disorder induced superconductivity in the rope which originates from the dynamics of the tubes. The latter enhance the intertube Josephson tunnelings which mitigate the suppression of the superconducting phase by disorder. However, for larger disorder amplitude, electronic localization prevails against intertube hopping leading to the suppression of superconductivity as found in other superconducting materials.

\section*{Acknowledgment}

We would like to acknowledge fruitful discussions with Pr. H. Bouchiat, Drs. M. Ferrier, S. Gu\'eron and K. Sasaki. We are grateful for Pr. H. Bouchiat for the critical reading of the manuscript. We warmly thank the staff of Laboratoire de
Physique des Solides \`a Orsay for kind hospitality.

\appendix
\setcounter{equation}{0}
\renewcommand{\theequation}%
{{\mbox{A}}.\arabic{equation}}
\section*{Appendix: Superconducting critical temperature}

By setting $\tilde{\epsilon}(T=T_c)=0$ in Eq.\ref{self} we obtain the following equation giving the superconducting critical temperature:

\begin{widetext}
\begin{eqnarray}
\ln\frac{Tc}{T_0}+gT\int^{\pi}_0\sin \theta d\theta \int^{2\pi}_0 d\varphi 
\int^{c}_0k^2 dk\int ^{W_c}_0 dW f(k,\theta,\varphi,W)=0.
\label{Tc}
\end{eqnarray}
\end{widetext}
The $f(k,\theta,\varphi,W)$ is given by:

\begin{widetext}
\begin{eqnarray}
f(k,\theta,\varphi,W)&=&\left\{ W^2+\frac{{\hbar}^2k^2}{2m^{\ast}}
+2\frac{J_0}{a_0}\left[1-\cos(kl_0\cos\theta)\right]\right.\nonumber\\
&+&2\frac{J_1}{a_0}\left[3-\cos(kd\sin\theta\cos\varphi)-\cos(kd\sin\theta\sin\varphi)-\cos\left(kd\sin\theta(\cos\varphi-\sin\varphi)\right)\right]\nonumber\\
&+&\left.2\frac{J_2}{a_0}\left[3-\cos\left(kd\sin\theta(2\cos\varphi-\sin\varphi)\right)-
\cos\left(kd\sin\theta(\cos\varphi+\sin\varphi)\right)-\cos\left(kd\sin\theta
(\cos\varphi-2\sin\varphi)\right)\right]\right\}^{-1}.
\end{eqnarray}
\end{widetext}

We introduced the dimensionless variable $W=\frac{\pi\hbar\omega}{8k_BT}$ and we set $g=8\pi\mu_0k_B\kappa^2e^2_0\xi^4_{\parallel}$, with $\lambda=0.6 \mu$m and $\xi_{\parallel}=\xi_0=0.1\mu$m \cite{Ferrier05}.\newline
We have adopted the no-cutoff limit for $c$ and $W_c$ ($c \rightarrow \infty$, $W_c \rightarrow \infty$), which means that all types of superconducting fluctuations, even with short wave lengths, are considered \cite{Puica}.

\end{document}